\documentclass[aps,prc,groupedaddress,showpacs,preprint]{revtex4-1}
\usepackage{graphicx}
\usepackage{colordvi}

\begin{document}

\title{Influence of differential elastic nucleon-nucleon cross section on stopping and collective flows in heavy-ion collisions at intermediate energies}
\author {Yongjia Wang$\, ^{1}$\footnote{wangyongjia@zjhu.edu.cn},
Chenchen Guo$\, ^{2,1}$, 
Qingfeng Li$\, ^{1}$\footnote{liqf@zjhu.edu.cn},
Zhuxia Li$\, ^{3}$,
Jun Su$\, ^{4}$,
and
Hongfei Zhang$\, ^{5}$
}

\affiliation{
1) School of Science, Huzhou University, Huzhou 313000, China \\
2) Shanghai Institute of Applied Physics, Chinese Academy of Sciences, Shanghai 201800, China\\
3) China Institute of Atomic Energy, Beijing 102413, China\\
4) Sino-French Institute of Nuclear Engineering and Technology, Sun Yat-sen University, Zhuhai 519082, China\\
5) School of Nuclear Science and Technology, Lanzhou University, Lanzhou 730000, China \\
 }
\date{\today}

\begin{abstract}
By considering three different Nucleon-Nucleon (NN) elastic differential cross sections: the Cugnon \emph{et al.} parameterized differential cross section [Nucl. Instrum. Methods Phys. Res., Sect. \textbf{B111}, 215 (1996)], and the differential cross section derived from the collision term of the self-consistent relativistic Boltzmann-Uehling-Uhlenbeck equation proposed by Mao \emph{et al.} [Z.\ Phys.\ A {\bf 347}, 173 (1994)], as well as the isotropic differential cross section, within the newly updated version of the ultrarelativistic quantum molecular dynamics (UrQMD)
model, the influence of the differential elastic NN cross section on various observables (e.g., nuclear stopping, both the rapidity and transverse-velocity dependence of the directed and elliptic flows) in Au+Au collisions at beam energies 150, 250, 400, and 800 MeV$/$nucleon is investigated. By comparing calculations with those three differential cross sections, it is found that the nuclear stopping power, the directed and elliptic flows are affected to some extent by the differential cross sections, and the impact of differential cross section on those observables becomes more visible as the beam energy increases. The effect on the elliptic flow difference
$v_{2}^{n}$-$v_{2}^{H}$ and ratio $v_{2}^{n}$/$v_{2}^{H}$ of neutrons versus hydrogen isotopes ($Z=1$), which have been used as sensitive observables for probing nuclear symmetry energy at high densities, is weak.

\end{abstract}

\pacs{21.65.Cd, 21.65.Mn, 25.70.-z}

\maketitle
\section{Introduction}

To understand the medium (density, isospin asymmetry, and temperature) dependence of the properties of nucleons and strong nucleon-nucleon interactions
is currently still one of the fundamental goals of nuclear physics. Particularly, the density dependence of the nuclear symmetry energy $E_{\rm sym}(\rho)$
which closely correlates to the isospin dependence of the strong interactions, has attracted considerable attention in
recent decades for its great importance for understanding the properties of nuclei far from stability as well as
neutron stars\cite{Baran:2004ih,Steiner:2004fi,BALi08,Lattimer:2006xb,DiToro:2010ku}. In recent several years, great efforts have been made to determine parameters (e.g., the coefficient $S_0=E_{\rm sym}(\rho_{0})$ and the slope $L=3\rho_{0}\left(\frac{\partial{E_{\rm sym}(\rho)}}{\partial\rho}\right)|_{\rho=\rho_{0}}$)
 of the symmetry energy at saturation density ($\rho_{0}$). So far, the picture of the nuclear symmetry energy around (below) $\rho_{0}$ becomes more and more clear but its value at high densities still has large uncertainties. (See, e.g., Refs.~\cite{Tsang:2012se,Lattimer:2012xj,RocaMaza:2012mh,Zhang:2013wna,brown,Danielewicz:2013upa,Fan:2014rha,Fan:2015vra,Zhang:2015ava,Wang:2015kof,Xiao:2009zza,Feng:2009am,Russotto:2011hq,Cozma:2013sja,Xie:2013np,wyj-sym}).

Heavy-ion collisions (HICs) provide a unique way to create nuclear matter with high density and isospin asymmetry ($\delta=\frac{\rho_{n}-\rho_{p}}{\rho_{n}+\rho_{p}}$, where $\rho_n$ and $\rho_p$ are the neutron and proton densities), but the created dense matter exists only for a very short time (typically several fm/c), and its properties cannot be measured directly in the laboratory. Thus transport models, which are used to describe the whole collision process and to deduce the properties of the intermediate stage from the assumed initial conditions and the final-state observables measured in the laboratory, are definitely needed. The most commonly employed
transport models when investigating HICs at low and intermediate energies are quantum molecular dynamics (QMD)\cite{Aichelin:1991xy} and Boltzmann (Vlasov)-Uehling-Uhlenbeck (BUU, VUU)\cite{Bertsch:1988ik} approaches.
At present, there are more than twenty improved versions of QMD-type and BUU-type models\cite{Xu:2016lue}. In both kinds of models, the mean field potential and nucleon-nucleon collisions are two essential parts\cite{Aichelin:1989pc,Hartnack:1997ez,Zhang:2010th}. The mean field potential in transport models has been carefully studied. For the collision part, the main inputs are integral and differential cross sections, the former determines the probability of two-body collisions while the latter determines the scattering angle in two-body collisions.
It should be stressed that the differential cross section is only used for the determination of the angular distribution in most versions of transport models but not for the corresponding integral total cross sections. In transport models, a parametrization of experimental differential cross section is usually used for convenience. For example, in the 1980s, the QMD model and BUU model used a differential cross section parameterized by Cugnon \emph{et al.}\cite{Cugnon:1980rb}, in which the isospin dependence has not been considered (neutron-neutron and neutron-proton scatterings were not distinguished). At present, more transport models use the new version parameterized by Cugnon \emph{et al.}\cite{Cugnon}, in which the isospin dependence has been considered. The Antisymmetrized molecular dynamics (AMD) model use another parameterized differential cross section proposed by Ono \emph{et al.}\cite{Ono:1993ac}. In the ultrarelativistic quantum molecular dynamics (UrQMD) model, an analytical expression for the differential cross section derived from the collision term of the self- consistent relativistic Boltzmann -Uehling-Uhlenbeck (RBUU) equation is used\cite{Mao:1994et,Mao:1996zz,Mao:1997gr,Bass98,Bleicher:1999xi}.

Certainly, the in-medium NN (differential) cross section can also be obtained by the relativistic (Dirac-)Brueckner approach or the Dirac-Brueckner-Hartree-Fock approaches\cite{bohnet,Li:1993rwa,Li:1993ef,Schulze:1997zz,Fuchs:2001fp,Zhang:2007zzs,Zhang:2010jf}. Unfortunately different approaches do not always give the same results. Thus, it is necessary to test these differential cross sections within transport models. Moreover,
there are many studies on the effect of the in-medium NN cross section on observables in HICs, but much less analysis has been made to investigate the effect of differential cross section. Particularly, large divergence has been shown when studying the high density behavior of the nuclear symmetry energy with transport model.
For example, the FOPI/LAND data\cite{Russotto:2011hq} on the elliptic flow ratio of neutrons with respect to protons or light complex particles were calculated by the UrQMD model with considering different stiffness of the nuclear symmetry energy, indicating a moderately soft to linear density dependence of the symmetry energy\cite{wyj-sym}. The result contrast with diverging results obtained from the comparisons of isospin-dependent Boltzmann-Uehling-Uhlenbeck or Lanzhou quantum molecular dynamics model calculations with the FOPI $\pi^-$/$\pi^+$ ratios from which both extremely soft and extremely stiff behaviors were extracted.
It is important to remark that meaningful constraints can be extracted from transport calculations only if the predictions of the model are not modified by uncontrolled model parameters. One of such parameters concerns the assumptions on the kinematics of the two-body collisions. Indeed, the angular dependence of the cross section can be modified by the in-medium effects, and this modification is model dependent (e.g. Refs\cite{Fuchs:2001fp,Zhang:2007zzs,DiazAlonso:1997xx}). This means that it is very important to assess whether a modification of the angular distribution has any influence upon some isospin-sensitive observables, particularly for the elliptic flow ratio of neutrons with respect to hydrogen isotopes which is supposed to be a good probe of the nuclear symmetry energy at high densities.

In this work, within the UrQMD model, we investigate the influence of the differential elastic NN cross section on nuclear stopping and collective flows in HICs at intermediate energies by considering three different differential cross sections. In the next section these differential NN elastic cross sections, and observables are introduced. In Sec. III, effects on stopping and collective flows of free protons from HICs at intermediate energies are shown and discussed. Finally, a summary is given in Sec. IV.

\section{Model descriptions and observables}
\label{sec:1}

With introducing the Skyrme potential-energy density functional in the mean-field potential part and an isospin-dependent minimum spanning tree algorithm (iso-MST) in present UrQMD code, the recent published experimental data can be reproduced quite well\cite{Wang:2013wca,wyj-sym}. In this work, the SV-mas08 and SV-sym34 interactions\cite{Dutra:2012mb,Klupfel:2008af}
are chosen which represent force with the incompressibility $K_0=233$ MeV and $K_0=234$ MeV, and the slope parameter of the nuclear symmetry energy $L=40$ MeV and $L=81$ MeV, respectively. The in-medium NN cross section and Pauli blocking treatments in the collision term
are taken in the same way as in our previous work in Ref.~\cite{Wang:2013wca}. For more details on the updated UrQMD model, we refer to Refs\cite{Li:2011zzp,Guo:2012aa,Wang:2012sy,Wang:2013wca,wyj-sym,Wang:2014aba}.
 The in-medium NN elastic cross section is suppressed compared to the free ones by considering a reduction factor in the transport model. Many experimental data in heavy-ion collisions at low and intermediate energies can be reproduced in this way, see, e.g., Ref.\cite{BALi08} for a review. However, the degree of suppression of the in-medium NN elastic cross section and its dependence on density, temperature, and momentum are still not well established. At energies above the pion production threshold, NN inelastic channels become more and more important, but the in-medium effects are still poorly studied. In the present work, for the NN inelastic channels, the experimental free-space cross sections are used. The energy up to 800 MeV/nucleon is chosen to show a larger effect of the differential elastic NN cross section on observables. It is higher than the pion production threshold but only observables related to nucleons are focused on. According to the estimation given in Ref.\cite{Bass:1995pj}, the probability for a nucleon to undergo inelastic scattering and to become a $\Delta$ is less than 10\%, thus the influence of inelastic channels on nucleonic observables that will be focused on in this work is small. Regarding the medium modification of the differential NN cross section, microscopic studies (see, e.g., Refs.\cite{Fuchs:2001fp,Zhang:2007zzs}) show that the presence of the nuclear medium tends to make the differential cross section more isotropic. On the contrary, when the screening effect from the nuclear medium was taken into account, the differential cross section becomes very forward-backward peaked at high density and energy (see, e.g., Refs\cite{DiazAlonso:1997xx}).
 To consider the uncertainty of the in-medium differential cross section, three commonly used differential NN cross sections in transport models are adopted and given as follows:

(i) The first one is the isotropic differential NN cross section named dcs\_iso. It means the cosine of the scattering angle between incident direction and scattered direction is randomly chosen between -1 and 1.

(ii) The second is the parametrization presented by Cugnon \emph{et al.}\cite{Cugnon} named dcs\_Cug. For proton-proton or neutron-neutron elastic scattering, the differential cross section can be calculated in the following way\cite{Cugnon}:
\begin{equation}
\frac{\mbox{d} \sigma_{\mbox{el}}^{pp}}{\mbox{d}t} =\frac{\mbox{d} \sigma_{\mbox{el}}^{nn}}{\mbox{d}t} \propto e^{B_{pp}t\quad},
\label{eqvd1}
\end{equation}
where $t$ and $s$ are the Mandelstam variables. $t=(\textbf{p}_1-\textbf{p}_3)^2=(\textbf{p}_2-\textbf{p}_4)^2$, related to the scattering angle, $s=(\textbf{p}_1+\textbf{p}_2)^2=(\textbf{p}_3+\textbf{p}_4)^2$ is also known as the square of the center-of-mass energy. Here $\textbf{p}_1$ and $\textbf{p}_2$ are the four-momenta of the incoming particles and $\textbf{p}_3$ and $\textbf{p}_4$ are the four-momenta of the outgoing particles in the two-body center-of-mass frame. For neutron-proton elastic scattering, the differential cross section can be calculated in the following way:
\begin{equation}
\frac{\mbox{d} \sigma_{\mbox{el}}^{np}}{\mbox{d}t} \propto e^{B_{np}t}+ae^{B_{np}u\quad},
\label{eqvd2}
\end{equation}
where $u$ is also the Mandelstam variable, $u=(\textbf{p}_1-\textbf{p}_4)^2=(\textbf{p}_2-\textbf{p}_3)^2$. Quantities $B_{pp}$,  $B_{np}$, and $a$ in Eqs.~\ref{eqvd1} and ~\ref{eqvd2} are a function of the center-of-mass energy and vary in different intervals of $\sqrt{s}$, or equivalently of $p_{lab}$, given as follows:

\begin{eqnarray}
B_{pp} = \left\{ \begin{array}{ll}
\frac{5.5p_{lab}^8}{7.7+p_{lab}^8} &\quad ;\quad  p_{lab}<2\\
5.334+0.67(p_{lab}-2) & \quad;\quad p_{lab}\geq2
\end{array} \right.
\end{eqnarray}

\begin{eqnarray}
B_{np} = \left\{ \begin{array}{ll}
0, &\quad ;\quad p_{lab}<0.225\\
16.53(p_{lab}-0.225)&\quad ;\quad0.225\leq p_{lab}<0.6\\
-1.63p_{lab}+7.16&\quad ;\quad 0.6\leq p_{lab}<1.6\\
B_{pp} &\quad ;\quad  p_{lab}\geq1.6
\end{array} \right.
\end{eqnarray}

\begin{eqnarray}
a = \left\{ \begin{array}{ll}
1 &\quad ;\quad  p_{lab}<0.8\\
\frac{0.64}{p_{lab}^2} & \quad;\quad p_{lab}\geq0.8
\end{array} \right.
\end{eqnarray}
Here $p_{lab}$ is the incident lab momentum in GeV$/$c.

\begin{figure}[htbp]
\centering
\includegraphics[angle=0,width=0.7\textwidth]{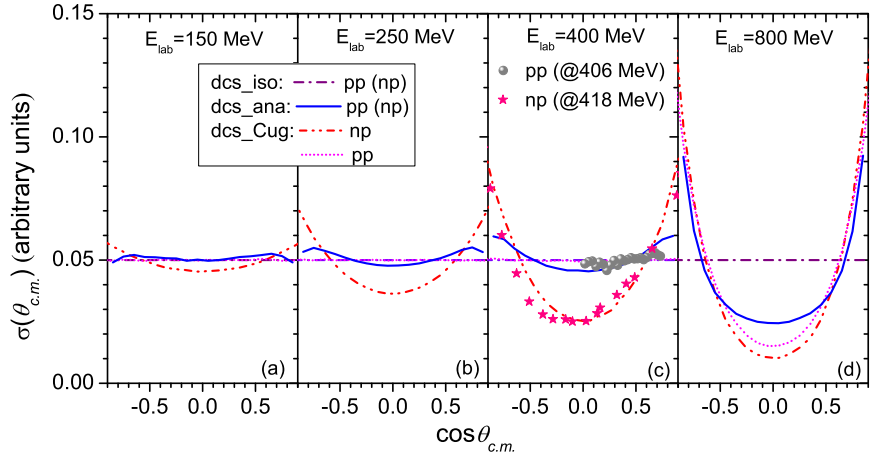}
\caption{\label{fig1}(Color online) The normalized differential cross sections vs the cosine of the center-of-mass scattering angle for neutron-proton and proton-proton (neutron-neutron) collisions at 150, 250, 400, and 800 MeV.
Results obtained by Cugnon \emph{et al.} (dcs\_Cug, dash-dot-dot line and dot line) and Mao \emph{et al.} (dcs\_ana, solid line) are compared to the isotropic differential cross section (dcs\_iso, dash-dot line). The experimental data for neutron-proton \cite{df-data} and proton-proton \cite{COSY} scatterings at near 400 MeV are shown.
 }
\end{figure}

(iii) The third in-medium differential NN elastic cross section is according to the analytical expressions given by Mao \emph{et al.}\cite{Bass98,Bleicher:1999xi}, named dcs\_ana. It reads:

\begin{equation}
  \sigma_{NN \rightarrow NN}(s,t) = \frac{1}{(2 \pi)^{2} s}
  \lbrack D(s,t)
  + E(s,t) + (s, t \longleftrightarrow u) \rbrack.
\label{angeq}  \end{equation}
Here $D(s,t)$ and $E(s,t)$ are the direct and exchange terms, their expressions can be found in Refs\cite{Bass98,Bleicher:1999xi}. In the UrQMD model, the Eq.~\ref{angeq} is used to determine the scattering angle for all hadron-hadron collisions under the assumption that the angular distributions
for all relevant two-body processes are similarly modified in a manner analogous to the NN elastic collision. It is worth stressing that the Cugnon parameterized differential cross section is isospin-dependent while the differential cross section used in the UrQMD model is isospin-independent. The in-medium NN cross sections in UrQMD are treated to be factorized as
the product of a medium correction factor and the free NN cross sections for which the isospin effect is considered.
As an example, we show in Fig.~\ref{fig1} the normalized differential cross sections as a function of the cosine of the center-of-mass scattering angle for neutron-proton and proton-proton (neutron-neutron) collisions at 150, 250, 400, and 800 MeV. Here the normalized differential cross section is the differential cross section divided by the integrated cross section. The normalized differential cross section for the dcs\_iso is equal to 0.05 because the center-of-mass scattering angle is divided into 20 bins. Since the differential cross section for proton-proton scattering at energy below 400 MeV is almost isotropic while for neutron-proton scattering is anisotropic (e.g., Refs\cite{df-data,COSY}), the experimental data for proton-proton \cite{COSY} and neutron-proton \cite{df-data} scatterings at near 400 MeV are shown to evaluate the degree of agreement among them. It can be seen that, the agreement between experimental data and the Cugnon parameterized values is good. We would like to note here again that even though the Cugnon parameterized formula can reproduced the experimental data (in free space) well, other assumptions on the differential cross section in the nuclear medium are still necessary for transport model, because the medium-modified differential cross section has not been well established. At low energy (i.e. 150 MeV), the dcs\_Cug and dcs\_ana are in accordance with the dcs\_iso, however at high energy (i.e. 400 and 800 MeV), larger differences among the three differential cross sections around $\theta_{c.m.}=90^\circ$ and $\theta_{c.m.}=0^\circ$ or $180^\circ$ can be found.

Nuclear stopping and the directed and elliptic flows are most commonly used observables in HICs at intermediate energies.
The directed and elliptic flow parameters $v_1$ and $v_2$ can be derived from the
Fourier expansion of the azimuthal distribution of detected particles as described in
Ref.~\cite{FOPI:2011aa}. We have
\begin{equation}
 \frac{dN}{u_{t}du_{t}dyd\phi} = v_0 [1 + 2v_1 \cos(\phi) + 2v_2
\cos(2\phi)] ,
\end{equation}
in which $\phi$ is the azimuthal angle of the emitted particle with
respect to the reaction plane, and $u_t=\beta_t
\gamma$ is the transverse component of the four-velocity
$u$=($\gamma$, $\bf{\beta}\gamma$), and rapidity $y=\frac{1}{2}\ln\frac{E+p_z}{E-p_z}$ where $p_z$ is the component of momentum along the beam axis.
The scaled units
$u_{t0}\equiv u_t/u_{1cm}$ and $y_0\equiv y/y_{1cm}$ are used throughout as done
in \cite{FOPI:2011aa}, and the subscript $1cm$ denotes
the incident projectile in the center-of-mass system.
The $v_1$ and $v_2$ are obtained from the following expressions:
  \begin{equation}
v_1\equiv \langle
cos(\phi)\rangle=\langle\frac{p_x}{p_t}\rangle;
v_2\equiv \langle cos(2\phi)\rangle=\langle\frac{p_x^2-p_y^2}{p_t^2}\rangle.
\label{eqv1}
\end{equation}
Here $p_t=\sqrt{p_x^2+p_y^2}$ is the
transverse momentum of emitted particles. The angle brackets in
Eq.~\ref{eqv1} indicate an average over all considered
particles from all events.

A possible measurement of the degree of stopping is the \emph{varxz}, the ratio of the variances of the transverse (usually refers to the \emph{x}-direction ) rapidity distribution over that of the longitudinal (the \emph{z}-direction) rapidity distributions, defined as\cite{FOPI:2010aa},
\begin{equation}
varxz=\frac{<y_{x}^2>}{<y_{z}^2>} . \label{eqvartl}
\end{equation}
Here
\begin{equation}
<y_{x,z}^2>=\frac{\sum(y^2_{x,z}N_{y_{x,z}})}{\sum
N_{y_{x,z}}}, \label{eqgm}
\end{equation}
where $<y_{x}^2>$ and $<y_{z}^2>$ are the
variances of the rapidity distributions of nucleons in the $x$ and
$z$ directions, respectively. $N_{y_{x}}$ and
$N_{y_{z}}$ denote the numbers of nucleons in each of the $y_x$
and $y_z$ rapidity bins.

\section{Results}
To show the effect of the NN elastic differential cross section on observables, $^{197}$Au+$^{197}$Au
collisions at beam energies 150, 250, 400 and 800 MeV$/$nucleon for centrality $0<b_0<0.55$ (the reduced impact parameter $b_0$ defined as
$b_0=b/b_{max}$, here $b_{max} = 1.15 (A_{P}^{1/3} +
A_{T}^{1/3})$~fm) are calculated.
The three-above mentioned differential NN elastic cross sections are considered, while other parts of the UrQMD model are treated in the same way.

\subsection{Influence on nuclear stopping}

\begin{figure}[htbp]
\centering
\includegraphics[angle=0,width=0.7\textwidth]{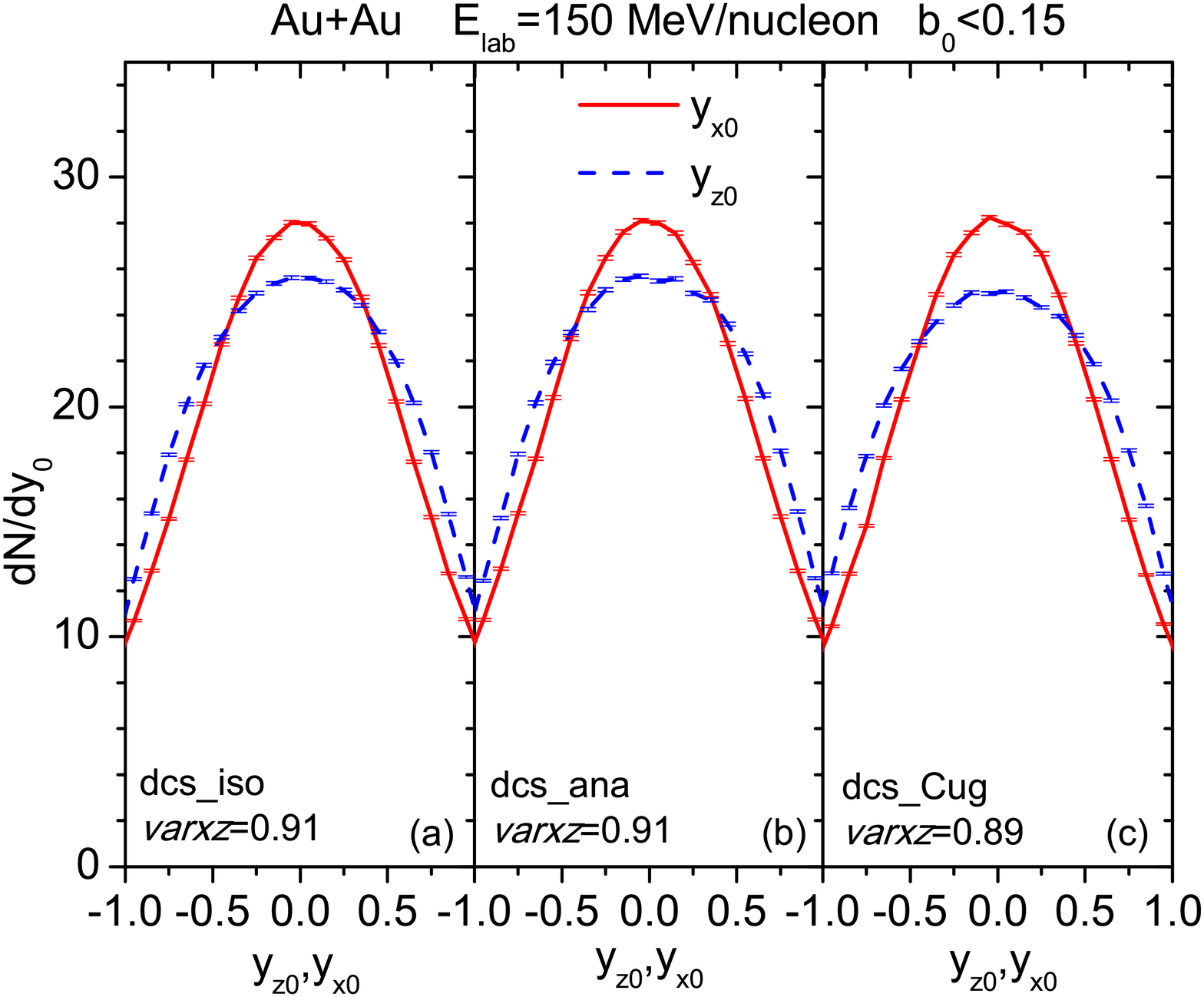}
\caption{\label{fig2}(Color online) Longitudinal and transverse rapidity distributions of free protons in central ($b_0$$\leq$0.15) $^{197}$Au+$^{197}$Au collisions at $E_{\rm lab}=150$~MeV$/$nucleon. Calculations preformed with dcs\_iso, dcs\_ana and dcs\_Cug are shown in the left (a), middle (b) and right (c) panels, respectively. The corresponding value of $varxz$ are also shown. }
\end{figure}

\begin{figure}[htbp]
\centering
\includegraphics[angle=0,width=0.7\textwidth]{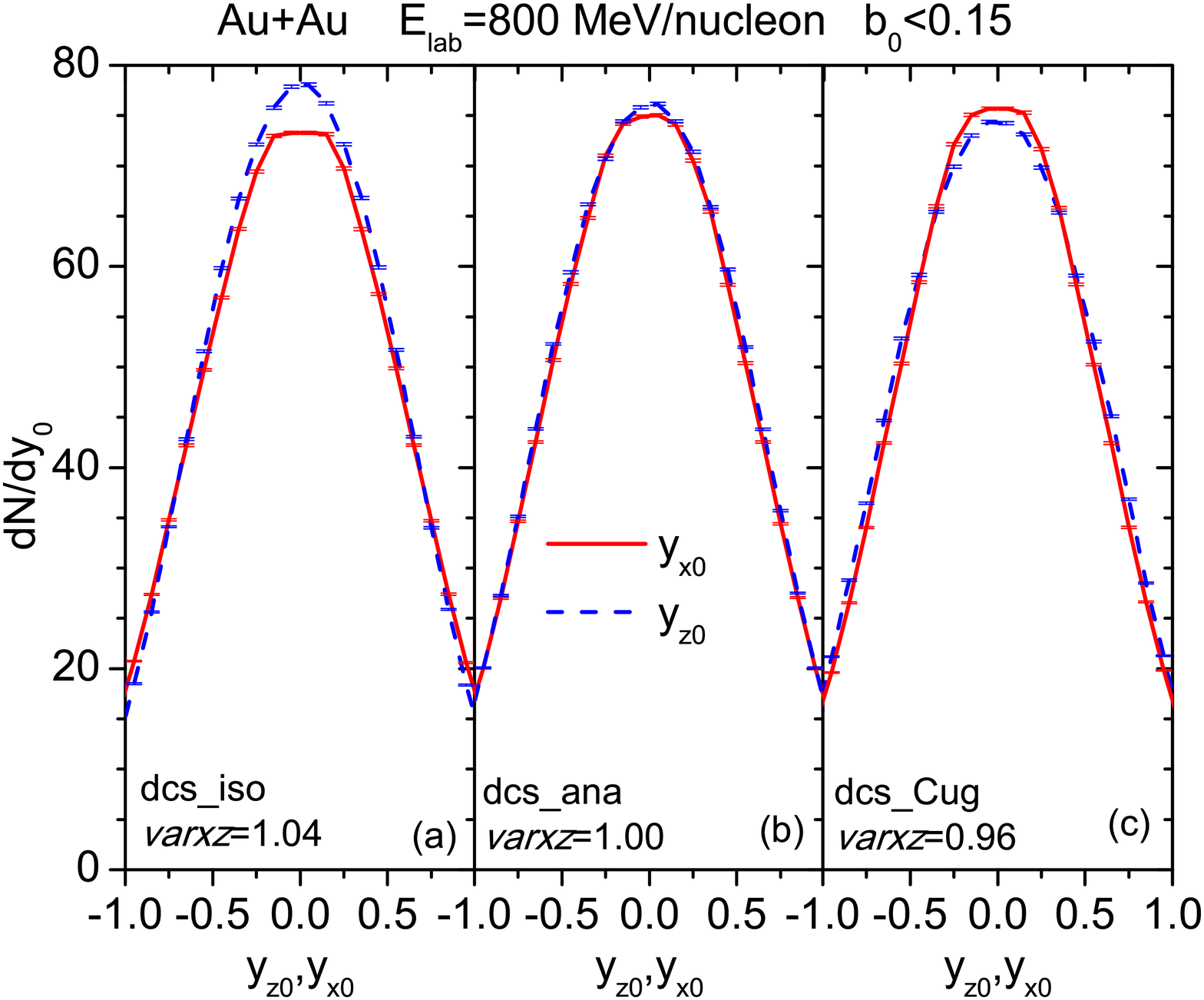}
\caption{\label{fig3}(Color online) Same as Fig.~\ref{fig2} but for $E_{\rm lab}=800$~MeV$/$nucleon.}
\end{figure}
First, the effects of the differential elastic NN cross section on the degree of stopping in central Au+Au collisions are shown in Figs.~\ref{fig2} and \ref{fig3}.
At low energy, i.e. at $E_{\rm lab}=150$~MeV$/$nucleon, both the longitudinal and transverse rapidity distributions obtained with the three differential cross sections are nearly the same, indicating that there is almost no difference in proton multiplicity. However, at high energy, i.e. at $E_{\rm lab}=800$~MeV$/$nucleon, as shown in Fig.~\ref{fig3}, the difference in $varxz$ can be clearly seen.
This can be understood from the fact that the angular distributions obtained from the three differential cross sections are quite the same at low center-of-mass energy, and that a distinction between them appears at high energy. A large difference in $varxz$ is expected to appear at high energy, but beam energies above 800~MeV$/$nucleon are not considered in this work because the NN inelastic channel will play an increasingly important role at higher energies. The value of $varxz$ calculated with dcs\_Cug is about 8\% less than that with dcs\_iso, the $varxz$ calculated with dcs\_ana is between the two others. It follows from the fact that a strong forward-backward peaking feature in the dcs\_Cug will make nucleons pass through each other more easily and then reduce the degree of nuclear stopping. In general, the change of differential cross section has a small influence on the degree of nuclear stopping and the influence increases with energy, consistent with the behavior of the differential cross sections changing with energy shown in Fig.~\ref{fig1}.

\subsection{Influence on collective flows}
\begin{figure}[htbp]
\centering
\includegraphics[angle=0,width=0.9\textwidth]{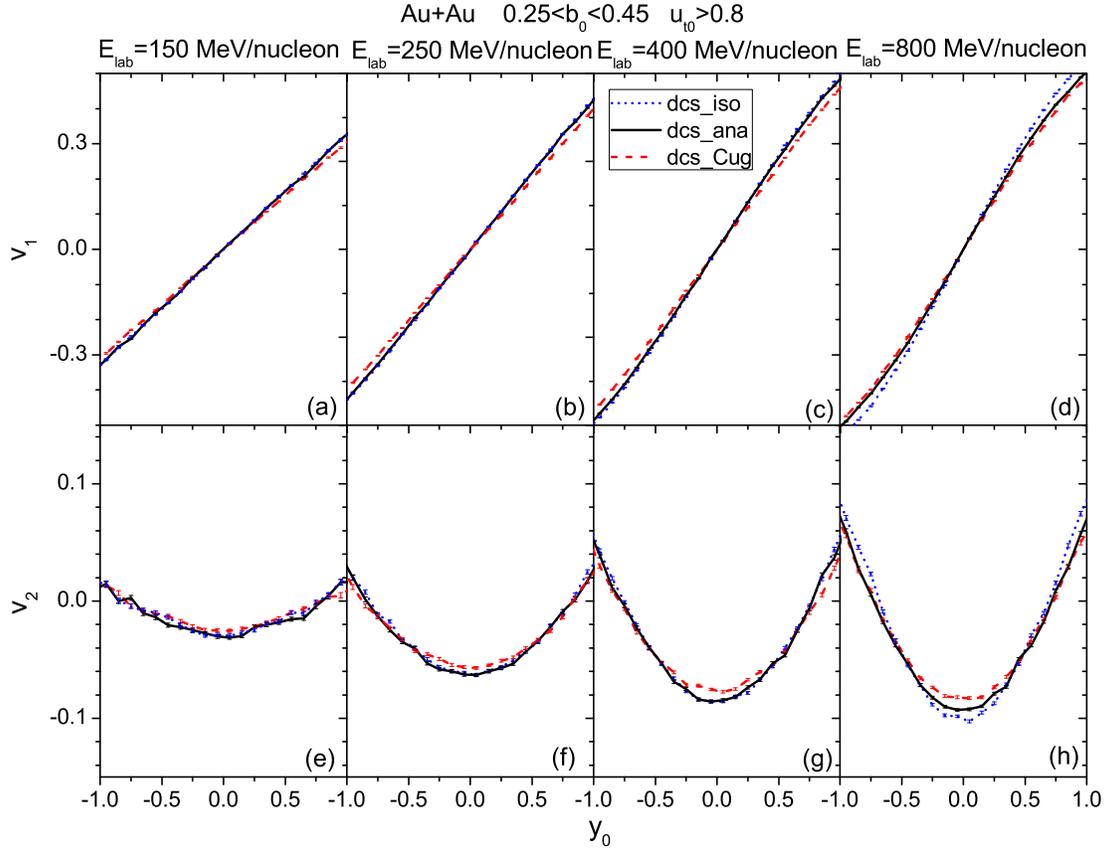}
\caption{\label{fig4}(Color online) Rapidity distribution of the directed flow $v_1$ (upper panels)
and elliptic flow $v_2$ (lower panels) of free protons from $^{197}$Au+$^{197}$Au collisions at 150, 250, 400, and 800 MeV$/$nucleon with centrality $0.25<b_0<0.45$ and $u_{t0}>0.8$.
}
\end{figure}

\begin{figure}[htbp]
\centering
\includegraphics[angle=0,width=0.9\textwidth]{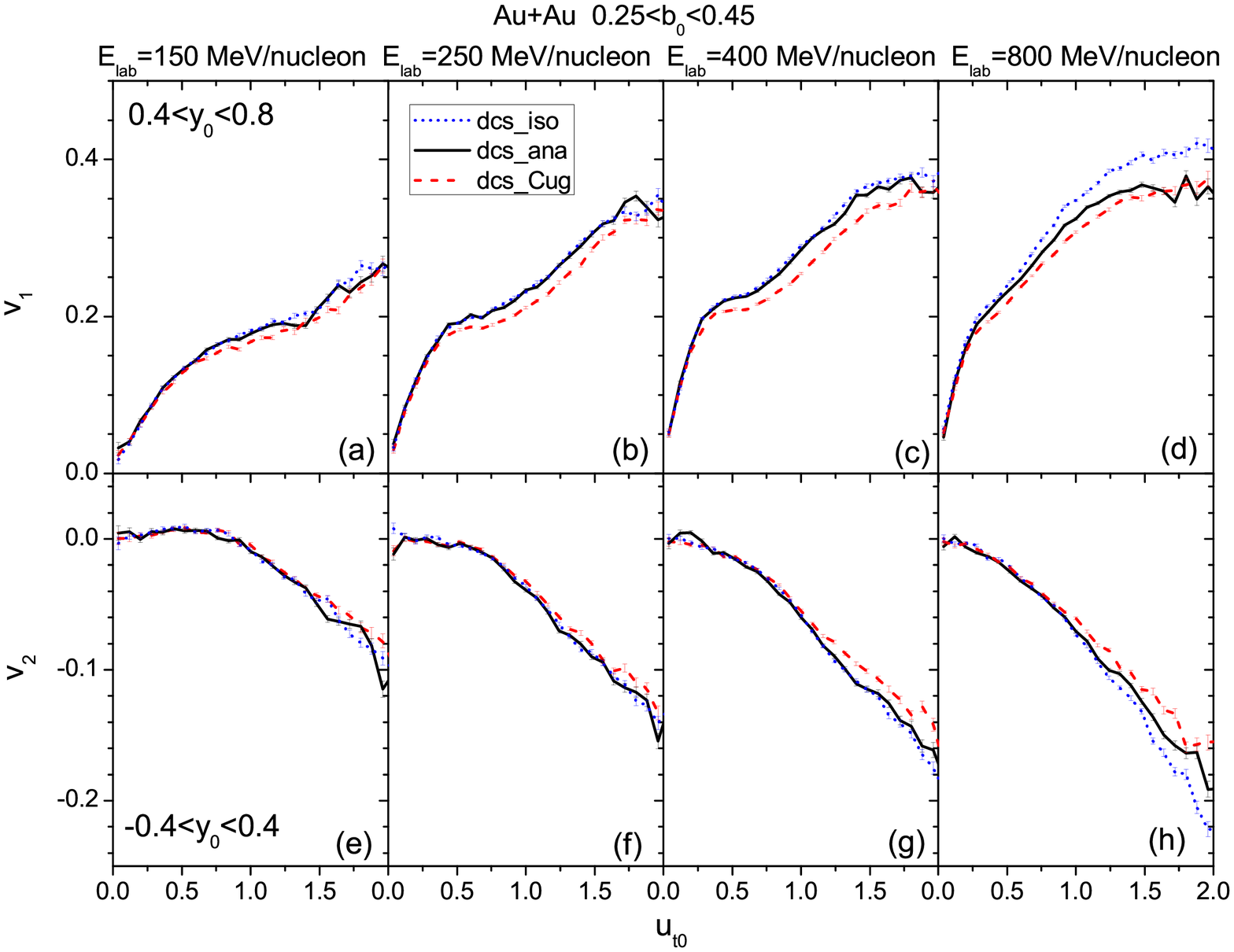}
\caption{\label{fig5}(Color online)  The directed flow $v_1$ (upper panels)
and elliptic flow $v_2$ (lower panels) of free protons as
a function of $u_{t0}$.
The $^{197}$Au+$^{197}$Au
collisions at the beam energy 150, 250, 400, and 800 MeV$/$nucleon with $0.25<b_0<0.45$
are considered. The rapidity cuts $0.4<y_0<0.8$ and $|y_0|<0.4$ are
chosen for $v_1$ and $v_2$, respectively.}
\end{figure}

Second, the influence of differential NN elastic cross section on the rapidity-dependent
directed and elliptic flows of free protons is shown in Fig.~\ref{fig4}. The $^{197}$Au+$^{197}$Au collisions at 150, 250, 400, and 800 MeV$/$nucleon with centrality $0.25<b_0<0.45$ are simulated by considering the three above-mentioned differential NN elastic cross sections. It can be seen from Fig.~\ref{fig4} that the directed flow $v_1$ and elliptic flow $v_2$ obtained with dcs\_ana and dcs\_iso approach each other quite closely at low beam energies, while both the $v_1$ and $v_2$ obtained with dcs\_Cug are slightly smaller than that obtained with the other two parametrizations. The difference between them steadily grows as the incident energy increases (cf. Fig.~\ref{fig4} (d) and (h)). The flow signal obtained with dcs\_Cug is the smallest, while that obtained with dcs\_iso is the largest in all three cases. The strong forward-backward peaking of the angular distribution of dcs\_Cug cause nucleons to be preferentially emitted along the initial direction (maintain the original momenta), thus reducing the flow signal. Moreover, the isotropic differential cross section will make nucleons undergo more rescattering and increase the blocking of the spectator matter, and apparently enhance the elliptic flow. We also find that, the slope of directed flow and the value of elliptic flow at midrapidity ($y_0$=0) obtained with dcs\_iso are about 15\% and 20\% larger than that obtained with dcs\_Cug. If one compares the influence of the differential cross section with that of the medium-modified total nucleon-nucleon cross section on the nuclear stopping and collective flow (e.g., Refs.\cite{wyj-sym,Wang:2013wca,Li:2011zzp}), it can be found that, in general, the influence of the medium-modified total nucleon-nucleon cross section which has not been well established is larger than that of the differential cross section.

To further investigate the influence of the differential NN elastic cross sections on the collective flow we show the parameters $v_1$ and $v_2$ as functions of $u_{t0}$ in Fig.~\ref{fig5} for beam energies 150, 250, 400, and 800 MeV$/$nucleon. The rapidity cuts are taken as $0.4<y_0<0.8$ for $v_1$ and $|y_0|<0.4$ for $v_2$. One sees that, as the energy increases, the difference between the results calculated with dcs\_iso, dfs\_cug and dcs\_ana increases. At 800 MeV$/$nucleon (i.e., Fig.~\ref{fig5} (d) and (h)), the influences of the differential cross sections on $v_1$ and $v_2$ become significant when $u_{t0}$ is larger than 1.0. It can actually be understood as follows: nucleons with high transverse momentum are usually emitted at early time and experience only a few scatterings and thus the pronounced differences among the three different differential cross sections around $\theta_{c.m.}=90^\circ$ and $\theta_{c.m.}=0^\circ$ or $180^\circ$ at high center-of-mass energy (see Fig.~\ref{fig1}) affect the flow signal. Thus the $v_1$ and $v_2$ of the high-transverse-momentum nucleons from high and more peripheral collisions seems sensitive to the angular distribution of the normalized differential cross section.

\subsection{Influence on isospin-sensitive observables}

\begin{figure}[htbp]
\centering
\includegraphics[angle=0,width=0.9\textwidth]{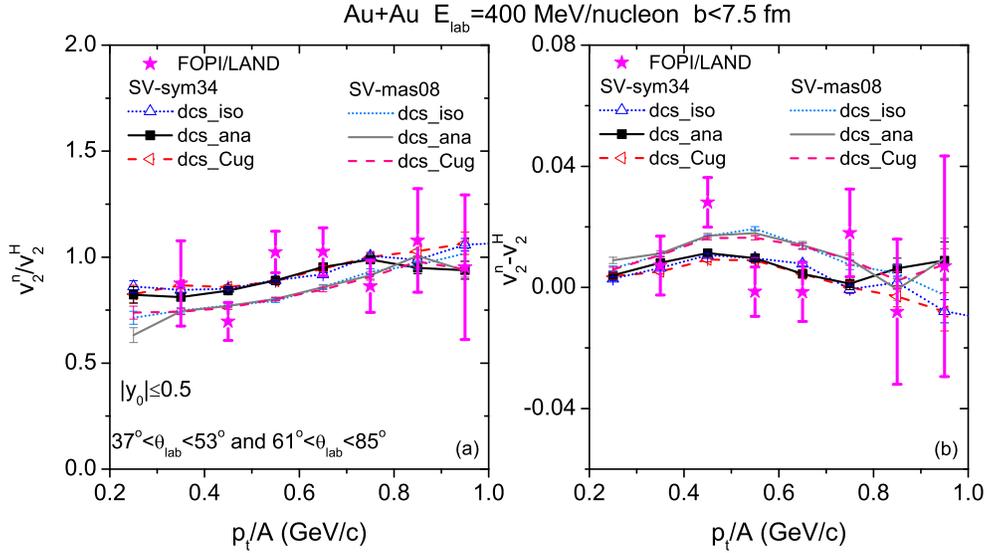}
\caption{\label{fig6}(Color online) Comparison of the elliptic flow ratio $v_{2}^{n}$/$v_{2}^{H}$ (left) and difference $v_{2}^{n}$-$v_{2}^{H}$ (right)
of free neutrons vs. hydrogen isotopes ($Z=1$) produced in central ($b<7.5$~fm) $^{197}$Au+$^{197}$Au collisions at $E_{\rm lab}=400$~MeV$/$nucleon
between UrQMD model and the FOPI/LAND data reported in
Ref.~\protect\cite{Russotto:2011hq}. The lines with symbols show calculations with SV-sym34, lines without symbols show calculations with SV-mas08. Cuts around midrapidity
$|y_0|\leq0.5$ and $\theta_{lab}$=$37^{\circ}$-$53^{\circ}$ and $61^{\circ}$-$85^{\circ}$ were employed.}
\end{figure}

It is therefore necessary to ascertain whether isospin-sensitive observables (e.g., the elliptic flow ratio $v_{2}^{n}$/$v_{2}^{H}$ and difference $v_{2}^{n}$-$v_{2}^{H}$ of free neutrons verse hydrogen isotopes) are affected by the differential NN elastic cross section. Fig.~\ref{fig6} shows a comparison of the elliptic flow ratio $v_{2}^{n}$/$v_{2}^{H}$ and difference $v_{2}^{n}$-$v_{2}^{H}$ for $^{197}$Au+$^{197}$Au collision at $E_{\rm lab}=400$~MeV$/$nucleon between the results of simulations and the FOPI/LAND data reported in Ref.~\cite{Russotto:2011hq}. SV-mas08 and SV-sym34 in combination with the three differential NN elastic cross sections are used. Clearly, both the $v_{2}^{n}$/$v_{2}^{H}$ and $v_{2}^{n}$-$v_{2}^{H}$ calculated with SV-mas08 and SV-sym34 are well separated and can be divided into two distinct groups. The influence of differential NN elastic cross section
 on the $v_{2}^{n}$/$v_{2}^{H}$ and $v_{2}^{n}$-$v_{2}^{H}$ is quite weak, especially in the low transverse momentum region in which the experimental error bars are relatively small.
It is similar to the weak effect of the total NN cross section on the elliptic flow ratio observed in Refs~\cite{Russotto:2011hq,wyj-sym}. The elliptic flow of neutrons
and hydrogen isotopes vary according to the total cross section, but the ratio between them does not change significantly.

\section{Summary}

In summary, by applying three frequently used differential nucleon-nucleon elastic cross sections (i.e., Cugnon \emph{et al.} parameterized differential cross section, Mao \emph{et al.} proposed differential cross section and the isotropic differential cross section) in the UrQMD model simulations, the influence of the differential cross section on nuclear stopping, the directed and elliptic flows of free protons produced in $^{197}$Au+$^{197}$Au collisions at $E_{\rm lab}$=150, 250, 400 and 800~MeV$/$nucleon is studied. It is found that both the nuclear stopping power and collective flows obtained by using the isotropic differential NN cross section are larger than those obtained by using the parametrization of Cugnon \emph{et al.} and the analytical expression given by Mao \emph{et al.}, for which a forward-backward peaking feature appears in the angular distribution. Moreover, the calculation results also show that the effect of the differential NN elastic cross section on observables increases with increasing energy because of the large divergence among the differential cross sections at high energies. At $E_{\rm lab}$=800~MeV$/$nucleon, the stopping power $varxz$, the slope of the directed flow and elliptic flow at midrapidity ($y_0$=0) obtained with the isotropic differential cross section are about 8\%, 15\% and 20\% larger than that obtained with the Cugnon parametrization, respectively. Thus, when obtaining constraints on the equation of state or the in-medium NN cross sections from heavy-ion reaction observables (such as nuclear stopping power and collective flows) in combination with transport model simulations, the uncertainty derived from the differential cross section should be considered. For the elliptic flow difference
$v_{2}^{n}$-$v_{2}^{H}$ and ratio $v_{2}^{n}$/$v_{2}^{H}$ of neutrons versus hydrogen isotopes ($Z=1$) which have been used as sensitive observables for probing the nuclear symmetry energy at high densities, our calculations show that the impact of the differential cross section on those observables is rather weak. This indicates that the constraint obtained on the nuclear symmetry energy from the elliptic flow ratio is not affected by systematic errors due to the uncertainty on the in-medium angular distribution of the elastic nucleon-nucleon cross section.

\begin{acknowledgments}
We thank Professor Wolfgang Trautmann for a careful reading of the
manuscript and valuable communications. One of us (Yongjia Wang) also wants to thank Yvonne Leifels and Arnaud Le F\`evre for their warm hospitality during his stay at GSI. The authors acknowledge support by the computing server C3S2 in Huzhou University. The work is supported in part by the National Natural
Science Foundation of China (Nos. 11505057, 11547312, 11375062, 11405278, 11175074, 11475262 and 11275052), by the Department of Education
of Zhejiang Province under Grant Y201533176, by the China Postdoctoral
Special Science Foundation funded project (No. 2016M591730), and the project sponsored by SRF for ROCS, SEM.
\end{acknowledgments}

\end{document}